\documentclass[10pt,paper,aps,prstab,twocolumn,amsmath,superscriptaddress,preprintnumbers,floatfix,nofootinbib,showpacs]{revtex4-2}

\usepackage{siunitx}
\usepackage{graphicx}
\usepackage[colorlinks]{hyperref}
\hypersetup{
	colorlinks=true, 
	linktoc=all,     
	linkcolor=blue,  
	citecolor=magenta,
	urlcolor=blue	
	}

\usepackage[dvipsnames]{xcolor}
\usepackage{tikz}

\usepackage{booktabs}
\usepackage{mathtools}
\usepackage{braket}
\usepackage[utf8]{inputenc}
\usepackage[T1]{fontenc}
\usepackage{natbib}
\usepackage{amssymb,latexsym}
\usepackage{wasysym}
\usepackage{soul}
\usepackage[super]{nth}
\usepackage{xfrac}
\usepackage{caption}
\usepackage{subcaption}

\begin{document}
	
\author{J. Slim}
\thanks{Present address: Deutsches Elektronen-Synchrotron, 22607 Hamburg, Germany}
\affiliation{III. Physikalisches Institut B, RWTH Aachen University, 52056 Aachen, Germany}

\author{F.\,Rathmann}
\email[Corresponding author: ]{frathmann@bnl.gov}\thanks{Present address: Brookhaven National Laboratory, Upton, NY 11973, USA}
\affiliation{Institut f\"ur Kernphysik, Forschungszentrum J\"ulich, 52425 J\"ulich, Germany}

\author{A.\,Andres}
\affiliation{III. Physikalisches Institut B, RWTH Aachen University, 52056 Aachen, Germany}
\affiliation{Institut f\"ur Kernphysik, Forschungszentrum J\"ulich, 52425 J\"ulich, Germany}

\author{V.\,Hejny}
\affiliation{Institut f\"ur Kernphysik, Forschungszentrum J\"ulich, 52425 J\"ulich, Germany}

\author{A.\,Nass}
\affiliation{Institut f\"ur Kernphysik, Forschungszentrum J\"ulich, 52425 J\"ulich, Germany}

\author{A.\,Kacharava}
\affiliation{Institut f\"ur Kernphysik, Forschungszentrum J\"ulich, 52425 J\"ulich, Germany}

\author{P.\,Lenisa}
\affiliation{University of Ferrara and Istituto Nazionale di Fisica Nucleare, 44100 Ferrara, Italy}

\author{N.N.\,Nikolaev}
\affiliation{L.D. Landau Institute for Theoretical Physics, 142432 	Chernogolovka, Russia}

\author{J.\,Pretz}
\affiliation{III. Physikalisches Institut B, RWTH Aachen University, 52056 Aachen, Germany}
\affiliation{Institut f\"ur Kernphysik, Forschungszentrum J\"ulich, 52425 J\"ulich, Germany}	

\author{A.\,Saleev}
\thanks{Present address: GSI Helmholtz Centre for Heavy Ion Research, 64291 Darmstadt, Germany}
\affiliation{University of Ferrara and Istituto Nazionale di Fisica Nucleare, 44100 Ferrara, Italy}

\author{V.\,Shmakova}
\thanks{Present address: Brookhaven National Laboratory, Upton, NY 11973, USA}
\affiliation{University of Ferrara and Istituto Nazionale di Fisica Nucleare, 44100 Ferrara, Italy}

\author{H.\,Soltner}
\affiliation{Zentralinstitut f\"ur Engineering, Elektronik und Analytik, Forschungszentrum J\"ulich, 52425 J\"ulich, Germany}

\author{F.\,Abusaif}
\thanks{Present  address: Karlsruhe Institute of Technology, 
	76344 Eggenstein-Leopoldshafen, Germany}
\affiliation{III. Physikalisches Institut B, RWTH Aachen University, 52056 Aachen, Germany}
\affiliation{Institut f\"ur Kernphysik, Forschungszentrum J\"ulich, 52425 J\"ulich, Germany}

\author{A.\,Aggarwal}
\affiliation{Marian Smoluchowski Institute of Physics, Jagiellonian University, 30348 Cracow, Poland}

\author{A.\,Aksentev}
\affiliation{Institute for Nuclear Research, Russian Academy of Sciences, 117312 Moscow, Russia}

\author{B.\,Alberdi}
\thanks{Present address: Humboldt-Universität zu Berlin, Institut für Physik, 
	12489 Berlin, Germany}
\affiliation{III. Physikalisches Institut B, RWTH Aachen University, 52056 Aachen, Germany}
\affiliation{Institut f\"ur Kernphysik, Forschungszentrum J\"ulich, 52425 J\"ulich, Germany}

\author{L.\,Barion}
\affiliation{University of Ferrara and Istituto Nazionale di Fisica Nucleare, 44100 Ferrara, Italy}

\author{I.\,Bekman}
\thanks{Present address: Zentralinstitut für Engineering, Elektronik und Analytik, Forschungszentrum Jülich, Jülich, Germany.}
\affiliation{Institut f\"ur Kernphysik, Forschungszentrum J\"ulich, 52425 J\"ulich, Germany}

\author{M.\,Bey\ss}
\affiliation{III. Physikalisches Institut B, RWTH Aachen University, 52056 Aachen, Germany}
\affiliation{Institut f\"ur Kernphysik, Forschungszentrum J\"ulich, 52425 J\"ulich, Germany}

\author{C.\,B\"ohme}
\affiliation{Institut f\"ur Kernphysik, Forschungszentrum J\"ulich, 52425 J\"ulich, Germany}

\author{B.\,Breitkreutz}
\thanks{Present address: GSI Helmholtz Centre for Heavy Ion Research, 64291 Darmstadt, Germany}
\affiliation{Institut f\"ur Kernphysik, Forschungszentrum J\"ulich, 52425 J\"ulich, Germany}

\author{N.\,Canale}
\affiliation{University of Ferrara and Istituto Nazionale di Fisica Nucleare, 44100 Ferrara, Italy}

\author{G.\,Ciullo}
\affiliation{University of Ferrara and Istituto Nazionale di Fisica Nucleare, 44100 Ferrara, Italy}

\author{S.\,Dymov}
\affiliation{University of Ferrara and Istituto Nazionale di Fisica Nucleare, 44100 Ferrara, Italy}

\author{N.-O.\,Fr\"ohlich}
\thanks{Present address: Deutsches Elektronen-Synchrotron, 22607 Hamburg, Germany}
\affiliation{Institut f\"ur Kernphysik, Forschungszentrum J\"ulich, 52425 J\"ulich, Germany}

\author{R.\,Gebel}
\affiliation{Institut f\"ur Kernphysik, Forschungszentrum J\"ulich, 52425 J\"ulich, Germany}

\author{M.\,Gaisser}
\affiliation{III. Physikalisches Institut B, RWTH Aachen University, 52056 Aachen, Germany}

\author{K.\,Grigoryev}
\thanks{Present address: GSI Helmholtz Centre for Heavy Ion Research, 64291 Darmstadt, Germany}
\affiliation{Institut f\"ur Kernphysik, Forschungszentrum J\"ulich, 52425 J\"ulich, Germany}

\author{D.\,Grzonka}
\affiliation{Institut f\"ur Kernphysik, Forschungszentrum J\"ulich, 52425 J\"ulich, Germany}

\author{J.\, Hetzel}
\affiliation{Institut f\"ur Kernphysik, Forschungszentrum J\"ulich, 52425 J\"ulich, Germany}

\author{O.\,Javakhishvili}
\affiliation{Department of Electrical and Computer Engineering, Agricultural University of Georgia, 0159 Tbilisi, Georgia}

\author{V.\,Kamerdzhiev}
\thanks{Present address: GSI Helmholtz Centre for Heavy Ion Research, 64291 Darmstadt, Germany}
\affiliation{Institut f\"ur Kernphysik, Forschungszentrum J\"ulich, 52425 J\"ulich, Germany}

\author{S.\,Karanth}
\affiliation{Marian Smoluchowski Institute of Physics, Jagiellonian University, 30348 Cracow, Poland}

\author{I.\,Keshelashvili}
\thanks{Present address: GSI Helmholtz Centre for Heavy Ion Research, 64291 Darmstadt, Germany}
\affiliation{Institut f\"ur Kernphysik, Forschungszentrum J\"ulich, 52425 J\"ulich, Germany}

\author{A.\,Kononov}
\affiliation{University of Ferrara and Istituto Nazionale di Fisica Nucleare, 44100 Ferrara, Italy}

\author{K.\,Laihem}
\thanks{Present address: GSI Helmholtz Centre for Heavy Ion Research, 64291 Darmstadt, Germany}
\affiliation{III. Physikalisches Institut B, RWTH Aachen University, 52056 Aachen, Germany}

\author{A.\,Lehrach}
\affiliation{III. Physikalisches Institut B, RWTH Aachen University, 52056 Aachen, Germany}
\affiliation{Institut f\"ur Kernphysik, Forschungszentrum J\"ulich, 52425 J\"ulich, Germany}


\author{N.\,Lomidze}
\affiliation{High Energy Physics Institute, Tbilisi State University, 0186 Tbilisi, Georgia}

\author{B.\,Lorentz}
\affiliation{GSI Helmholtzzentrum für Schwerionenforschung, 64291 Darmstadt, Germany}

\author{G.\,Macharashvili}
\affiliation{High Energy Physics Institute, Tbilisi State University, 0186 Tbilisi, Georgia}

\author{A.\,Magiera}
\affiliation{Marian Smoluchowski Institute of Physics, Jagiellonian University, 30348 Cracow, Poland}

\author{D.\,Mchedlishvili}
\affiliation{High Energy Physics Institute, Tbilisi State University, 0186 Tbilisi, Georgia}

\author{A.\,Melnikov}
\affiliation{Institute for Nuclear Research, Russian Academy of Sciences, 117312 Moscow, Russia}
	
\author{F.\,Müller}
\affiliation{III. Physikalisches Institut B, RWTH Aachen University, 52056 Aachen, Germany}
\affiliation{Institut f\"ur Kernphysik, Forschungszentrum J\"ulich, 52425 J\"ulich, Germany}

\author{A.\,Pesce}
\affiliation{Institut f\"ur Kernphysik, Forschungszentrum J\"ulich, 52425 J\"ulich, Germany}

\author{V.\,Poncza}
\affiliation{Institut f\"ur Kernphysik, Forschungszentrum J\"ulich, 52425 J\"ulich, Germany}

\author{D.\,Prasuhn}
\affiliation{Institut f\"ur Kernphysik, Forschungszentrum J\"ulich, 52425 J\"ulich, Germany}

\author{D.\,Shergelashvili}
\affiliation{High Energy Physics Institute, Tbilisi State University, 0186 Tbilisi, Georgia}	

\author{N.\,Shurkhno}
\thanks{Present address: GSI Helmholtz Centre for Heavy Ion Research, 64291 Darmstadt, Germany}
\affiliation{Institut f\"ur Kernphysik, Forschungszentrum J\"ulich, 52425 J\"ulich, Germany}

\author{S.\,Siddique}
\thanks{Present address: GSI Helmholtz Centre for Heavy Ion Research, 64291 Darmstadt, Germany}
\affiliation{III. Physikalisches Institut B, RWTH Aachen University, 52056 Aachen, Germany}
\affiliation{Institut f\"ur Kernphysik, Forschungszentrum J\"ulich, 52425 J\"ulich, Germany}

\author{A.\,Silenko}
\affiliation{Bogoliubov Laboratory of Theoretical Physics, Joint Institute for Nuclear Research, 141980 Dubna, Russia}

\author{S.\,Stassen}
\affiliation{Institut f\"ur Kernphysik, Forschungszentrum J\"ulich, 52425 J\"ulich, Germany}

\author{E.J.\,Stephenson}		
\affiliation{Indiana University, Department of Physics, Bloomington, Indiana 47405, USA}

\author{H.\,Ströher}
\affiliation{Institut f\"ur Kernphysik, Forschungszentrum J\"ulich, 52425 J\"ulich, Germany}

\author{M.\,Tabidze}
\affiliation{High Energy Physics Institute, Tbilisi State University, 0186 Tbilisi, Georgia}

\author{G.\,Tagliente}
\affiliation{Istituto Nazionale di Fisica Nucleare sez.\ Bari, 70125 Bari, Italy}

\author{Y.\,Valdau}
\thanks{Present address: GSI Helmholtz Centre for Heavy Ion Research, 64291 Darmstadt, Germany}
\affiliation{Institut f\"ur Kernphysik, Forschungszentrum J\"ulich, 52425 J\"ulich, Germany}

\author{M.\,Vitz}
\affiliation{III. Physikalisches Institut B, RWTH Aachen University, 52056 Aachen, Germany}
\affiliation{Institut f\"ur Kernphysik, Forschungszentrum J\"ulich, 52425 J\"ulich, Germany}

\author{T.\,Wagner}
\thanks{Present address: GSI Helmholtz Centre for Heavy Ion Research, 64291 Darmstadt, Germany}
\affiliation{III. Physikalisches Institut B, RWTH Aachen University, 52056 Aachen, Germany}
\affiliation{Institut f\"ur Kernphysik, Forschungszentrum J\"ulich, 52425 J\"ulich, Germany}

\author{A.\,Wirzba}
\affiliation{Institut f\"ur Kernphysik, Forschungszentrum J\"ulich,  52425 J\"ulich, Germany}
\affiliation{Institute for Advanced Simulation, Forschungszentrum J\"ulich, 52425 J\"ulich, Germany}

\author{A.\,Wro\'{n}ska}
\affiliation{Marian Smoluchowski Institute of Physics, Jagiellonian University, 30348 Cracow, Poland}

\author{P.\,W\"ustner}
\affiliation{Zentralinstitut f\"ur Engineering, Elektronik und Analytik, Forschungszentrum J\"ulich, 52425 J\"ulich, Germany}

\author{M.\,\.{Z}urek}
\thanks{Present address: Argonne National Laboratory, Lemont, IL 60439, USA}
\affiliation{Institut f\"ur Kernphysik, Forschungszentrum J\"ulich, 52425 J\"ulich, Germany}

\collaboration{JEDI collaboration}
	
	
\date {\today}
	

\title{
Pilot bunch and co-magnetometry of polarized particles stored in a ring
}
\begin{abstract}
In  polarization experiments at storage rings, one of the challenges is to maintain the spin-resonance condition of a radio-frequency spin rotator with the spin-precessions of the orbiting particles. Time-dependent variations of the magnetic fields of ring elements  lead to unwanted variations of the spin precession frequency. We report here on a solution to this problem by shielding (or masking) one of the bunches stored in the ring from the high-frequency fields of the spin rotator, so that the masked \textit{pilot} bunch acts as a co-magnetometer for the other \textit{signal} bunch, tracking fluctuations in the ring on a time scale of about one second. While the new method was developed primarily for searches of electric dipole moments of charged particles, it may have far-reaching implications for future spin physics facilities, such as the EIC and NICA.
\end{abstract}
\maketitle
	

Controlled radio-frequency (RF) driven spin rotations, in particular the spin flip, are indispensable for nuclear physics experiments with polarized particles (see \textit{e.g.}, \cite{PhysRevC.58.658,pd:IUCF}, for reviews, see\,\cite{SYLee, Yokoya}). Extensive spin physics experiments in storage rings are either performed\,\cite{morozov2003first, Meyer:2007zzb, PhysRevSTAB.18.020101, ABDULKHALEK2022122447} or prepared, in particular to search for physics beyond the Standard Model (BSM)\,\cite{srEDM,AbusaifCYR,HybridEDMring,Snowmass:EDM,OkunMillistrong,PrentkiMillistrong,LeeMillistrong,NikolaevPrecessingD}, where it is essential to maintain the exact spin resonance condition for a long time to allow a large number of spin flips during the continuous operation of an RF spin rotator.

One example for precision studies addressing BSM physics is the search for the  electric dipole moment (EDM) of charged particles, which requires to accumulate the EDM driven spin rotation signal during a long spin-coherence time\,\cite{srEDM,AbusaifCYR,HybridEDMring,Snowmass:EDM}. In principle, the co-magnetometry can be provided by the oscillating \textit{horizontal} polarization of the stored beam interacting with an internal polarimeter target which results in an up-down asymmetry that oscillates with the spin-precession frequency. A Fourier analysis of the time-stamped events in the polarimeter allows one to determine the oscillation frequency and thus the spin precession frequency with $\approx \num{e-10}$  accuracy within a time window of \SI{100}{s}\,(see Ref.\,\cite{PhysRevLett.115.094801,JEDIspintune2} for details). 

Our studies, however, revealed a non-negligible variation of the idle spin-precession frequency in the ring on the level of about \num{e-8} from one fill to another and during each fill (see\,\cite[Fig.\,4]{PhysRevLett.115.094801}). In order to compensate for possible systematic biases and to maintain the spin-resonance condition, \textit{continuous} co-magnetometry is required to  provide feedback to the RF spin flipper in terms of frequency and phase information, as discussed in detail in\,\cite{JEDIphase}. However, when the spins are closely aligned along the vertical axis in the machine during single or multiple spin flips (SF), the horizontal polarization component disappears, rendering the control of the spin-precession frequency impossible.

In this communication by the JEDI\footnote{J\"ulich  Electric Dipole moment Investigations \url{http://collaborations.fz-juelich.de/ikp/jedi/}} collaboration, we report about a solution to the co-magnetometry problem based on the so-called \textit{pilot bunch} approach, for which we have successfully executed a proof-of-principle experiment at the Cooler Synchrotron (COSY) at Forschungszentrum Jülich. The demonstration was performed with polarized deuterons stored in the ring and made use of a radio-frequency Wien filter (WF) as a Lorentz force-free spin-flipper\,\cite{Slim2016116,SlimWFCircuit}. The basic idea is to store multiple bunches of particles whose spins precess around the vertical guiding field of the ring dipole magnets. Subsequently, the RF Wien filter is used in a special mode in which it acts as a spin-flipper on all but one of the bunches, turning off once per beam revolution for a specified time interval when the bunch acting as a co-magnetometer, the \textit{pilot bunch}, passes through the spin-flipper.

It should be noted that our approach to co-magnetometry in EDM searches for charged particles using storage rings differs significantly from the advanced mercury ($^{199}\text{Hg}$) co-magnetometer used in EDM searches with ultra-cold neutrons. There, the neutrons and the mercury atoms are different species, both essentially at rest, with the mercury atoms probing the fields to which the neutrons are exposed\,\cite{Altarev:1981zp,abel2020measurement}. In our approach, however, the pilot bunch acts as a co-magnetometer, probing the electromagnetic environment in which the other bunches orbiting in the ring are moving.

We first describe the experimental setup 
and operation of the RF Wien filter in gate mode  with two stored beam bunches, where the fast RF switches, developed in collaboration with the company Barthel\,\footnote{Barthel HF–Technik GmbH, 52072 Aachen, Germany \url{https://barthel-hf.com}}, were included in the driving circuit. Then we proceed to the description of the proof of principle of the pilot-bunch approach to the co-magnetometry. While the present work focused primarily on co-magnetometry, a much broader range of RF gating applications is conceivable. In particular, gating the high frequency of spin flippers opens up the possibility of reversing the polarization of selected bunches \textit{during} each store, which allows us to organize successive bunches with alternating vertical beam polarizations, leading to a reduction of systematic errors in spin asymmetry experiments at future colliders such as  NICA (Nuclotron-based Ion Collider fAcility in Dubna)\,\cite{Kekelidze:2016hhw} and EIC (Electron-Ion Collider in Brookhaven)\,\cite{ABDULKHALEK2022122447}.


The basic demonstration of the pilot-bunch approach was carried out with deuterons at a flattop momentum of $\SI{970}{MeV/c}$. In each cycle (fill) the vector polarized deuterons were injected, bunched in two packages, each containing about \num{e9} particles, electron-cooled for about a minute at \SI{76}{MeV} down to a momentum spread of $\Delta p/p \approx \num{e-4}$, and then accelerated to flattop. The beam is stochastically extracted on flattop onto a carbon block target at the  JEDI polarimeter\,\cite{Mueller:2020}, which is used to monitor the horizontal,  $p_x$, and   vertical,  $p_y$,  polarization components of the beam. Details about the machine timing sequence and beam and machine parameters are given in the Supplementary Material\,\cite[Sec.\,I]{Supplementary}.

Prior to the experiments, the initially vertical spins of the stored deuterons were rotated once into the horizontal plane by an LC-resonant RF solenoid\,\cite[Sec.\,7.7.3]{AbusaifCYR}, operated at a fixed frequency. The frequency needed to accomplish that is determined by observing the vanishing of $p_y$ in the polarimeter. In the further course of the experiment, the spin-precession frequency $f_\text{s}$ of the in-plane polarization, determined only rather roughly in this way, was used as the starting frequency for the operation of the RF Wien filter to ensure the resonance condition  $f_\text{WF}=f_\text{s} \pm  K f_\text{rev}$, where $K \in \mathbb{Z}$ is the sideband and $f_\text{rev}$ the beam revolution frequency. In the present experiment, the Wien filter was run at $K= -1$. In an ideal storage ring, free of magnetic imperfections, the spin-precession frequency $f_\text{s}=G\gamma f_\text{rev}$, $G$ the magnetic anomaly and $\gamma$ the relativistic factor of the particle, but in practice the magnetic ring imperfection effects might be substantial\,\cite{SpinTuneMapping}. It should be noted that for the proof-of-principle experiment described here, satisfying the  resonance condition \textit{exactly} is not mandatory (see discussion in ref.\,\cite[Sec.\,III\,A]{Nikolaev:2023srx}).


The experiment starts with  two back-to-back bunches orbiting in the machine with their spins aligned along the vertical axis, perpendicular to the ring plane. After electron cooling is switched off at $t_\text{cyc} = 77 \,\text{s}$, the periphery of the beam is brought into interaction with the carbon polarimeter target by stochastic heating using a stripline. 

\begin{figure}[htb]
	\begin{subfigure}[b]{\columnwidth}
		\includegraphics[width=\textwidth]{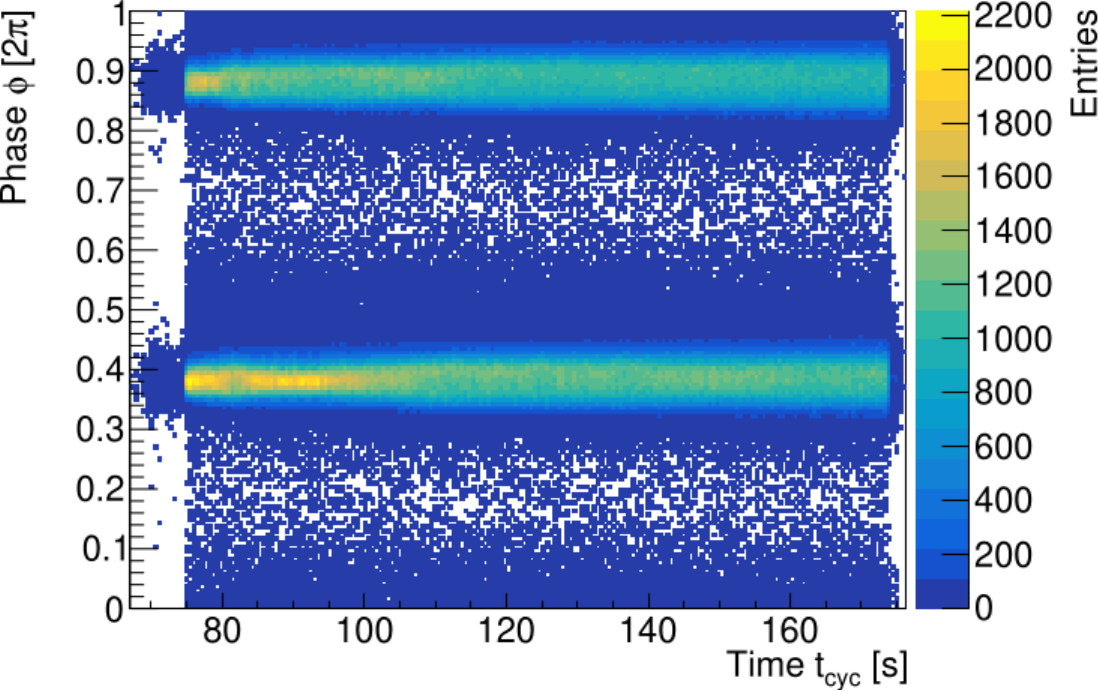}
		\caption{The pilot bunch is located near a phase of $\phi_\text{p} \simeq \SI{2.4}{rad}$ and the signal bunch near $\phi_\text{s} \simeq \SI{5.6}{rad}$ ($2\pi$ denotes the ring circumference).}
		\label{fig:TwoBunches:a}
	\end{subfigure}
	\begin{subfigure}[b]{\columnwidth}
		\includegraphics[width=\textwidth]{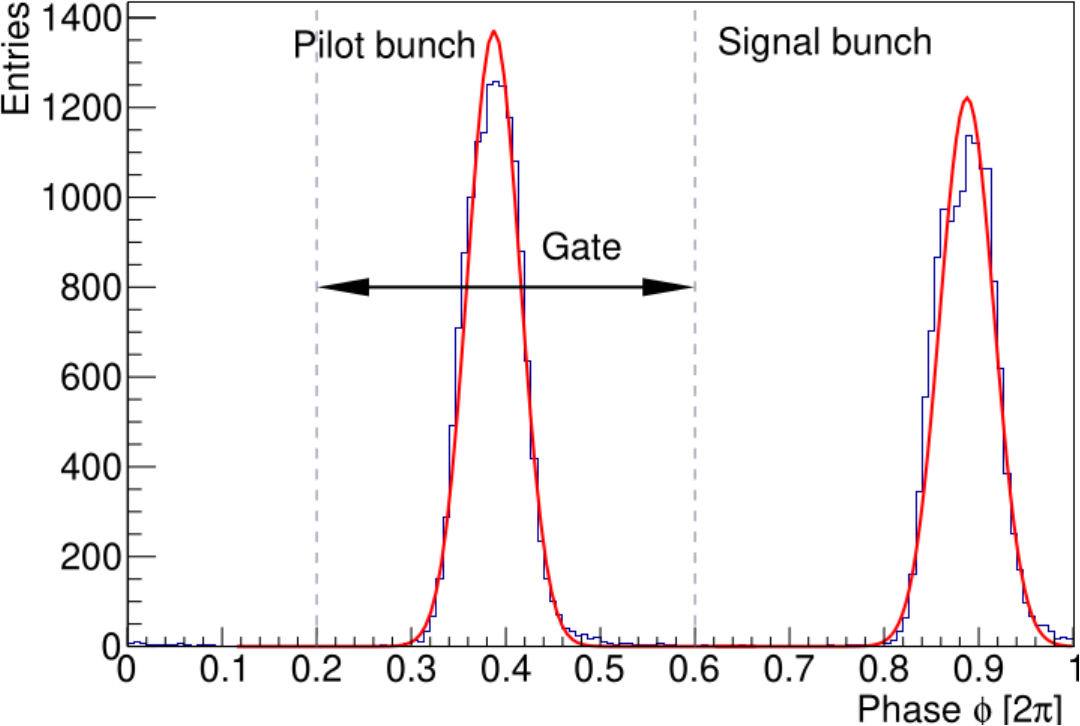}
		\caption{Entries (counts) recorded in the detector system during a time interval of \SI{1}{s}, plotted as a function of revolution phase, reflect the longitudinal beam profiles of the two back-to-back deuteron bunches in the ring at a cycle time of $t_\text{cyc} = \SI{122}{s}$. The total number of entries in the spectrum corresponds to about \num{38000} events. The vertical lines indicate the width of the gate that is used to mask the pilot bunch from the RF of the Wien filter.}
		\label{fig:TwoBunches:b}
	\end{subfigure}
	\caption{Time distributions of events recorded from interactions of the beam with the carbon polarimeter target. Panel (a) shows a 2D plot of the evolution of two bunches stored in the ring as function of time.  Panel (b) shows the beam distributions at $t_\text{cyc} = \SI{122}{s}$ as a function of the revolution phase, given by Eq.\,(\ref{eq:Time-to-Phase}). Beam widths and bunch separations from fits with Gaussian at three instances during the cycle are summarized  in the Supplementary Material\,\cite[Sec.\,II]{Supplementary}.}
	\label{fig:TwoBunches}
\end{figure}

The time distribution of the events recorded in the polarimeter is mapped into the revolution phase $\phi$, given by
\begin{equation}
\phi = 2\pi \left[ f_\text{rev} t_\text{cyc} - \text{int}( f_\text{rev}t_\text{cyc}) \right] \in [0,2\pi]\,.
\label{eq:Time-to-Phase}
\end{equation}
where $2\pi$ corresponds to the ring circumference. The time evolution of the two bunches, pilot (p) and signal (s), is plotted as a function of cycle time $t_\text{cyc}$ in Fig.\,\ref{fig:TwoBunches:a}. 
The bunch length is increasing due to emittance growth. An example of the longitudinal beam profile of both bunches near the mid point of the cycle at $t_\text{cyc} = \SI{122}{s}$ is depicted in Fig.\,\ref{fig:TwoBunches:b} as a function of the revolution phase $\phi$.  The gate width is well sufficient to fully shield the pilot bunch from the RF field of the Wien filter, the details of the pilot and signal bunch parameters are reported  in the Supplementary Material\,\cite[Sec.\,II]{Supplementary}.


The pilot bunch is gated out by fast RF switches 
in the input and output lines of the Wien filter. The function of these switches is to render the RF of the WF invisible to one of the bunches orbiting in the ring, while the signal bunches are  subjected to RF-driven multiple spin flips. The details of the switch operation are described in the Supplementary Material\,\cite[Sec.\,III]{Supplementary}.

The function of the pilot bunch as a co-magnetometer derives from its insensitivity to the operation of the Wien filter, so that its polarization continues to idly precess in the ring plane, thus \textit{continuously} collecting information about the spin precession. This information is used to correct the frequency of the Wien filter in a feedback system to maintain the  resonance condition.  A similar feedback system, stabilizing the idle spin-precession frequency by reducing and enhancing the beam's revolution frequency had been applied earlier\,\cite{JEDIphase}. 

\begin{figure}[tb]
	\includegraphics[width=\columnwidth]{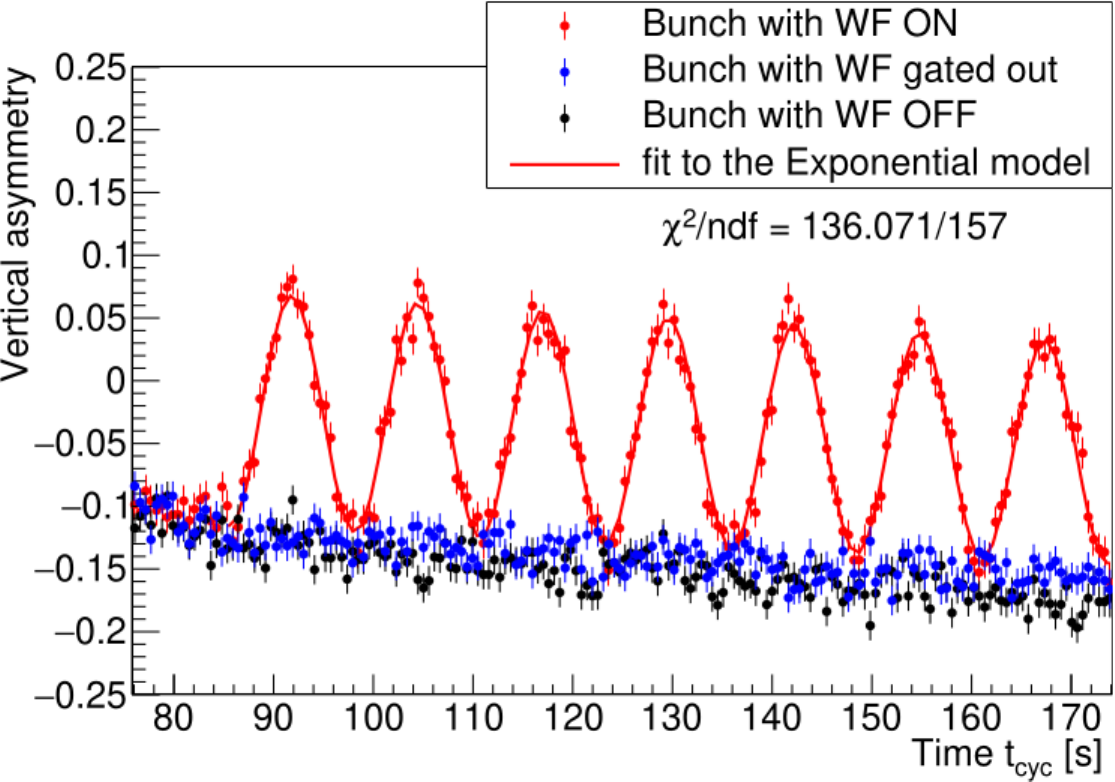}
	\caption{\label{fig:Three-polarizations} The measured left-right asymmetry  induced  by the RF Wien filter in the polarimeter is presented as a vertical oscillation of the beam polarization for a cycle with two bunches stored  in the machine, as depicted in Fig.\,\ref{fig:TwoBunches:b}. (The dC analyzing power is not yet applied,)  The red points indicate the vertical polarization asymmetry when at $t_0 = \SI{85.55}{s}$ the RF Wien filter is switched ON (signal bunch) with an  additional $\pm 2 \sigma$ cut on the signal bunch distribution. The blue points reflect the case for the pilot bunch, \textit{i.e.}, when the RF of the Wien filter is gated out, as indicated in Figs.\,\ref{fig:TwoBunches:b} and \cite[Fig.\,S2]{Supplementary}. The black points indicate the situation when, during a different cycle, the Wien filter is completely switched OFF. The red line indicates a fit with Eq.\,(\ref{eq:frequency-fit}), using events from within the $\pm 2\sigma_\text{s}$ boundary of the signal bunch distribution, and the results obtained are given in the Supplementary Material\,\cite[Sec.\,IV\,A]{Supplementary}.}
\end{figure}

The experimental result of the test of the pilot bunch principle is illustrated in Fig.\,\ref{fig:Three-polarizations}. Recording the time stamp of the interactions in the detectors of the polarimeter allows for a concurrent 
measurement of the left-right asymmetries caused by the pilot and signal bunches. It should be noted that for the experimental proof of the pilot-bunch technique aimed at here, only asymmetries need to be taken into account; calibrated polarizations are not required. As the target intercepts the periphery of the beam, the off-centered interactions may induce a finite offset of the measured asymmetries and also exhibit a slow time-dependence caused by the enhanced beam heating to maintain a constant count rate, but these are arguably independent of the beam polarization and would not affect the principal distinction between the pilot and signal bunches.

The signal  bunch (red symbols) exhibits the expected multiple continuous spin flips (SF). In striking contrast, the asymmetry measured for the pilot bunch (blue symbols) shows no oscillation signal at the spin-flip frequency $f_\text{SF}$ and perfectly matches that measured in a cycle where the Wien filter was off (black symbols), with the caveat that here we are forced to compare data recorded in different fills.

In the phenomenological analysis, the observed asymmetries shall be described by  function
\begin{equation}
A(t) = a (t-t_0) + b
       + c  \exp\left(- \Gamma (t-t_0)\right) \cos\left[2\pi f_\text{SF} (t-t_0)\right]\, ,
\label{eq:frequency-fit}
\end{equation}
where an allowance is made for the spin decoherence caused damping in terms of a time constant $\tau =1/\Gamma$. The oscillations of the beam-spin asymmetries of the signal bunch are shown in Fig.\,\ref{fig:Three-polarizations} and were fitted using Eq.\,(\ref{eq:frequency-fit}), which yields the spin-flip amplitude $c$ and the frequency $f_\text{SF}$. The results of the spin flip analysis are discussed in the Supplementary Material\,\cite[Sec.\,IV]{Supplementary}.

To some extent, synchrotron oscillations in the stored beam may contribute to synchrotron-amplitude dependent detuning of the spin oscillations of the  central and head \& tail regions of the beam bunches and can result in off-resonance behavior. These aspects are discussed in great detail in ref.\,\cite{Nikolaev:2023srx}. We were able to investigate whether the head \& tail regions of the signal bunch, which are populated by particles with larger synchrotron amplitudes, have different oscillation frequencies than the central regions. The results, presented in the Supplementary Material\,\cite[Sec.\,IV\,B]{Supplementary}, show that no differences in the spin-flip amplitudes and spin-flip frequencies $f_\text{SF}$ for the different regions were found within the errors.

Multiple spin flips are often described in terms of the efficiency $\epsilon_\text{flip}$, {\it i.e.,} by the ratio of polarizations after and before a single spin flip (see \textit{e.g.}, refs.\,\cite{morozov2003first, Meyer:2007zzb, PhysRevSTAB.18.020101}). In terms of our parametrization in Eq.\,(\ref{eq:frequency-fit}), the spin-flip efficiency in our experiment can be expressed via
\begin{equation}
\epsilon_\text{SF} = 1 -\frac{\Gamma}{2f_\text{SF}}\,.
\end{equation}
In the present experiment, the observed attenuation of the polarization amplitude proved to be very weak, and the resulting single spin-flip efficiency is essentially compatible with unity (see Supplementary Material\,\cite[Sec.\,IV\,A]{Supplementary}.)
 
Based on the spin flip frequency $f_\text{SF}$ from the signal bunch, we can quantify the gating quality by determining the oscillation amplitude $c_\text{p}$ of the asymmetries for the \textit{pilot bunch} by fitting with the same function $A(t)$, but this time with fixed $t_0$ and $f_\text{SF}$. As the amplitude of oscillation approaches zero, the attenuation parameter $\Gamma$ is indeterminate and was fixed at the value found for the signal bunch. As described in the Supplementary Material\,\cite[Sec.\,IV\,C]{Supplementary}, in terms of the oscillation amplitudes for the pilot and signal bunches $c_\text{p}$ and $c_\text{s}$, we obtain a gating efficiency of
\begin{equation}
\epsilon_\text{gate} = 1 - \frac{c_\text{p}}{c_\text{s}} = \num{0.9921} \pm \num{0.0135}\,,
\end{equation} 
compatible with unity, which indicates that the pilot-bunch approach has performed remarkably well and that our fast prototype RF switches were operating very close to perfection.



We demonstrated the feasibility of the pilot bunch based co-magnetometry for storage ring experiments which is imperative for high-precision spin experiments. The pilot-bunch technique has been primarily proposed in the first place for precision spin experiments that involve testing of fundamental symmetries, such as searches for the parity- and time-reversal-invariance violating permanent EDMs of charged particles\,\cite{srEDM,AbusaifCYR}, but it may find other applications in the field of spin physics at storage rings. As an example, we mention in this context the search for millistrong $CP$ violation\,\cite{OkunMillistrong,PrentkiMillistrong,LeeMillistrong} via the measurement of time reversal-odd spin asymmetries in interactions of tensor polarized deuterons with polarized protons, where the in-plane precessing spins of deuterons would give rise to a $T$-odd asymmetry that oscillates with twice the spin-precession frequency, free of systematics \cite{NikolaevPrecessingD}.

As a related example of self-co-magnetometry, consider the search for axions using polarized particles in storage rings as axion antennas\,\cite{JEDI:2022hxa}: if there is experimental indication of an axion resonance at certain frequencies, one can use the beam itself as a magnetometer to stabilize the spin-precession frequency at the suspected axion field oscillation frequency to enhance the axion signal.
 
We would also like to emphasize that \textit{gating out} the pilot bunch can alternatively be viewed as \textit{gating in} the signal bunch. Besides providing a solution to the problem of co-magnetometry in precision experiments, this opens up new possibilities for spin physics experiments at multi-bunch accelerators such as EIC and NICA, making use of Lorentz force-free spin manipulators, as highlighted here. Instead of injecting polarized bunches with a predefined alternating polarization pattern  into the collider, one can invert the vertical polarization \textit{on flattop} by  selectively gating individual bunches or groups of bunches and thus reduce the systematic errors, \textit{e.g.}, in double-polarized deep inelastic scattering. 

\begin{acknowledgments}
We would like to thank the COSY crew for their support in setting up the COSY accelerator for the experiment. The work presented here has been performed in the framework of the JEDI collaboration and was supported by an ERC Advanced Grant of the European Union (proposal No.\,694340: Search for electric dipole moments using storage rings) and by the Shota Rustaveli National Science Foundation of the Republic of Georgia (SRNSFG Grant No.\,DI-18-298: High precision polarimetry for charged particle EDM searches in storage rings). This research is part of a project that has received funding from the European Union’s Horizon 2020 research and innovation program under grant agreement STRONG-2020 No.\,824093. The work of A.\,Aksentev, A.\,Melnikov and N.\,Nikolaev on the topic was supported by the Russian Science Foundation (Grant No.\,22-42-04419).
\end{acknowledgments}

\bibliographystyle{apsrev4-2}
\bibliography{PilotBunch_18.06.2023.bib}

\end{document}


	
\author{J. Slim}
\thanks{Present address: Deutsches Elektronen-Synchrotron, 22607 Hamburg, Germany}
\affiliation{III. Physikalisches Institut B, RWTH Aachen University, 52056 Aachen, Germany}

\author{F.\,Rathmann}
\email[Corresponding author: ]{frathmann@bnl.gov}
\thanks{Present address: Brookhaven National Laboratory, Upton, NY 11973, USA}
\affiliation{Institut f\"ur Kernphysik, Forschungszentrum J\"ulich, 52425 J\"ulich, Germany}

\author{A.\,Andres}
\affiliation{III. Physikalisches Institut B, RWTH Aachen University, 52056 Aachen, Germany}
\affiliation{Institut f\"ur Kernphysik, Forschungszentrum J\"ulich, 52425 J\"ulich, Germany}

\author{V.\,Hejny}
\affiliation{Institut f\"ur Kernphysik, Forschungszentrum J\"ulich, 52425 J\"ulich, Germany}

\author{A.\,Nass}
\affiliation{Institut f\"ur Kernphysik, Forschungszentrum J\"ulich, 52425 J\"ulich, Germany}

\author{A.\,Kacharava}
\affiliation{Institut f\"ur Kernphysik, Forschungszentrum J\"ulich, 52425 J\"ulich, Germany}

\author{P.\,Lenisa}
\affiliation{University of Ferrara and Istituto Nazionale di Fisica Nucleare, 44100 Ferrara, Italy}

\author{N.N.\,Nikolaev}
\affiliation{L.D. Landau Institute for Theoretical Physics, 142432 	Chernogolovka, Russia}

\author{J.\,Pretz}
\affiliation{III. Physikalisches Institut B, RWTH Aachen University, 52056 Aachen, Germany}
\affiliation{Institut f\"ur Kernphysik, Forschungszentrum J\"ulich, 52425 J\"ulich, Germany}	

\author{A.\,Saleev}
\thanks{Present address: GSI Helmholtz Centre for Heavy Ion Research, 64291 Darmstadt, Germany}
\affiliation{University of Ferrara and Istituto Nazionale di Fisica Nucleare, 44100 Ferrara, Italy}

\author{V.\,Shmakova}
\thanks{Present address: Brookhaven National Laboratory, Upton, NY 11973, USA}
\affiliation{University of Ferrara and Istituto Nazionale di Fisica Nucleare, 44100 Ferrara, Italy}

\author{H.\,Soltner}
\affiliation{Zentralinstitut f\"ur Engineering, Elektronik und Analytik, Forschungszentrum J\"ulich, 52425 J\"ulich, Germany}

\author{F.\,Abusaif}
\thanks{Present  address: Karlsruhe Institute of Technology, 76344 Eggenstein-Leopoldshafen, Germany}
\affiliation{III. Physikalisches Institut B, RWTH Aachen University, 52056 Aachen, Germany}
\affiliation{Institut f\"ur Kernphysik, Forschungszentrum J\"ulich, 52425 J\"ulich, Germany}

\author{A.\,Aggarwal}
\affiliation{Marian Smoluchowski Institute of Physics, Jagiellonian University, 30348 Cracow, Poland}

\author{A.\,Aksentev}
\affiliation{Institute for Nuclear Research, Russian Academy of Sciences, 117312 Moscow, Russia}

\author{B.\,Alberdi}
\thanks{Present address: Humboldt-Universität zu Berlin, Institut für Physik, 12489 Berlin, Germany}
\affiliation{III. Physikalisches Institut B, RWTH Aachen University, 52056 Aachen, Germany}
\affiliation{Institut f\"ur Kernphysik, Forschungszentrum J\"ulich, 52425 J\"ulich, Germany}

\author{L.\,Barion}
\affiliation{University of Ferrara and Istituto Nazionale di Fisica Nucleare, 44100 Ferrara, Italy}

\author{I.\,Bekman}
\thanks{Present address: Zentralinstitut für Engineering, Elektronik und Analytik, Forschungszentrum Jülich, Jülich, Germany.}
\affiliation{Institut f\"ur Kernphysik, Forschungszentrum J\"ulich, 52425 J\"ulich, Germany}

\author{M.\,Bey\ss}
\affiliation{III. Physikalisches Institut B, RWTH Aachen University, 52056 Aachen, Germany}
\affiliation{Institut f\"ur Kernphysik, Forschungszentrum J\"ulich, 52425 J\"ulich, Germany}

\author{C.\,B\"ohme}
\affiliation{Institut f\"ur Kernphysik, Forschungszentrum J\"ulich, 52425 J\"ulich, Germany}

\author{B.\,Breitkreutz}
\thanks{Present address: GSI Helmholtz Centre for Heavy Ion Research, 64291 Darmstadt, Germany}
\affiliation{Institut f\"ur Kernphysik, Forschungszentrum J\"ulich, 52425 J\"ulich, Germany}

\author{N.\,Canale}
\affiliation{University of Ferrara and Istituto Nazionale di Fisica Nucleare, 44100 Ferrara, Italy}

\author{G.\,Ciullo}
\affiliation{University of Ferrara and Istituto Nazionale di Fisica Nucleare, 44100 Ferrara, Italy}

\author{S.\,Dymov}
\affiliation{University of Ferrara and Istituto Nazionale di Fisica Nucleare, 44100 Ferrara, Italy}

\author{N.-O.\,Fr\"ohlich}
\thanks{Present address: Deutsches Elektronen-Synchrotron, 22607 Hamburg, Germany}
\affiliation{Institut f\"ur Kernphysik, Forschungszentrum J\"ulich, 52425 J\"ulich, Germany}

\author{R.\,Gebel}
\affiliation{Institut f\"ur Kernphysik, Forschungszentrum J\"ulich, 52425 J\"ulich, Germany}

\author{M.\,Gaisser}
\affiliation{III. Physikalisches Institut B, RWTH Aachen University, 52056 Aachen, Germany}

\author{K.\,Grigoryev}
\thanks{Present address: GSI Helmholtz Centre for Heavy Ion Research, 64291 Darmstadt, Germany}
\affiliation{Institut f\"ur Kernphysik, Forschungszentrum J\"ulich, 52425 J\"ulich, Germany}

\author{D.\,Grzonka}
\affiliation{Institut f\"ur Kernphysik, Forschungszentrum J\"ulich, 52425 J\"ulich, Germany}

\author{J.\, Hetzel}
\affiliation{Institut f\"ur Kernphysik, Forschungszentrum J\"ulich, 52425 J\"ulich, Germany}

\author{O.\,Javakhishvili}
\affiliation{Department of Electrical and Computer Engineering, Agricultural University of Georgia, 0159 Tbilisi, Georgia}

\author{V.\,Kamerdzhiev}
\thanks{Present address: GSI Helmholtz Centre for Heavy Ion Research, 64291 Darmstadt, Germany}
\affiliation{Institut f\"ur Kernphysik, Forschungszentrum J\"ulich, 52425 J\"ulich, Germany}

\author{S.\,Karanth}
\affiliation{Marian Smoluchowski Institute of Physics, Jagiellonian University, 30348 Cracow, Poland}

\author{I.\,Keshelashvili}
\thanks{Present address: GSI Helmholtz Centre for Heavy Ion Research, 64291 Darmstadt, Germany}
\affiliation{Institut f\"ur Kernphysik, Forschungszentrum J\"ulich, 52425 J\"ulich, Germany}

\author{A.\,Kononov}
\affiliation{University of Ferrara and Istituto Nazionale di Fisica Nucleare, 44100 Ferrara, Italy}

\author{K.\,Laihem}
\thanks{Present address: GSI Helmholtz Centre for Heavy Ion Research, 64291 Darmstadt, Germany}
\affiliation{III. Physikalisches Institut B, RWTH Aachen University, 52056 Aachen, Germany}

\author{A.\,Lehrach}
\affiliation{III. Physikalisches Institut B, RWTH Aachen University, 52056 Aachen, Germany}
\affiliation{Institut f\"ur Kernphysik, Forschungszentrum J\"ulich, 52425 J\"ulich, Germany}


\author{N.\,Lomidze}
\affiliation{High Energy Physics Institute, Tbilisi State University, 0186 Tbilisi, Georgia}

\author{B.\,Lorentz}
\affiliation{GSI Helmholtzzentrum für Schwerionenforschung, 64291 Darmstadt, Germany}

\author{G.\,Macharashvili}
\affiliation{High Energy Physics Institute, Tbilisi State University, 0186 Tbilisi, Georgia}

\author{A.\,Magiera}
\affiliation{Marian Smoluchowski Institute of Physics, Jagiellonian University, 30348 Cracow, Poland}

\author{D.\,Mchedlishvili}
\affiliation{High Energy Physics Institute, Tbilisi State University, 0186 Tbilisi, Georgia}

\author{A.\,Melnikov}
\affiliation{Institute for Nuclear Research, Russian Academy of Sciences, 117312 Moscow, Russia}

\author{F.\,Müller}
\affiliation{III. Physikalisches Institut B, RWTH Aachen University, 52056 Aachen, Germany}
\affiliation{Institut f\"ur Kernphysik, Forschungszentrum J\"ulich, 52425 J\"ulich, Germany}

\author{A.\,Pesce}
\affiliation{Institut f\"ur Kernphysik, Forschungszentrum J\"ulich, 52425 J\"ulich, Germany}

\author{V.\,Poncza}
\affiliation{Institut f\"ur Kernphysik, Forschungszentrum J\"ulich, 52425 J\"ulich, Germany}

\author{D.\,Prasuhn}
\affiliation{Institut f\"ur Kernphysik, Forschungszentrum J\"ulich, 52425 J\"ulich, Germany}

\author{D.\,Shergelashvili}
\affiliation{High Energy Physics Institute, Tbilisi State University, 0186 Tbilisi, Georgia}	

\author{N.\,Shurkhno}
\thanks{Present address: GSI Helmholtz Centre for Heavy Ion Research, 64291 Darmstadt, Germany}
\affiliation{Institut f\"ur Kernphysik, Forschungszentrum J\"ulich, 52425 J\"ulich, Germany}

\author{S.\,Siddique}
\thanks{Present address: GSI Helmholtz Centre for Heavy Ion Research, 64291 Darmstadt, Germany}
\affiliation{III. Physikalisches Institut B, RWTH Aachen University, 52056 Aachen, Germany}
\affiliation{Institut f\"ur Kernphysik, Forschungszentrum J\"ulich, 52425 J\"ulich, Germany}

\author{A.\,Silenko}
\affiliation{Bogoliubov Laboratory of Theoretical Physics, Joint Institute for Nuclear Research, 141980 Dubna, Russia}

\author{S.\,Stassen}
\affiliation{Institut f\"ur Kernphysik, Forschungszentrum J\"ulich, 52425 J\"ulich, Germany}

\author{E.J.\,Stephenson}		
\affiliation{Indiana University, Department of Physics, Bloomington, Indiana 47405, USA}

\author{H.\,Ströher}
\affiliation{Institut f\"ur Kernphysik, Forschungszentrum J\"ulich, 52425 J\"ulich, Germany}

\author{M.\,Tabidze}
\affiliation{High Energy Physics Institute, Tbilisi State University, 0186 Tbilisi, Georgia}

\author{G.\,Tagliente}
\affiliation{Istituto Nazionale di Fisica Nucleare sez.\ Bari, 70125 Bari, Italy}

\author{Y.\,Valdau}
\thanks{Present address: GSI Helmholtz Centre for Heavy Ion Research, 64291 Darmstadt, Germany}
\affiliation{Institut f\"ur Kernphysik, Forschungszentrum J\"ulich, 52425 J\"ulich, Germany}

\author{M.\,Vitz}
\affiliation{III. Physikalisches Institut B, RWTH Aachen University, 52056 Aachen, Germany}
\affiliation{Institut f\"ur Kernphysik, Forschungszentrum J\"ulich, 52425 J\"ulich, Germany}

\author{T.\,Wagner}
\thanks{Present address: GSI Helmholtz Centre for Heavy Ion Research, 64291 Darmstadt, Germany}
\affiliation{III. Physikalisches Institut B, RWTH Aachen University, 52056 Aachen, Germany}
\affiliation{Institut f\"ur Kernphysik, Forschungszentrum J\"ulich, 52425 J\"ulich, Germany}

\author{A.\,Wirzba}
\affiliation{Institut f\"ur Kernphysik, Forschungszentrum J\"ulich,  52425 J\"ulich, Germany}
\affiliation{Institute for Advanced Simulation, Forschungszentrum J\"ulich, 52425 J\"ulich, Germany}

\author{A.\,Wro\'{n}ska}
\affiliation{Marian Smoluchowski Institute of Physics, Jagiellonian University, 30348 Cracow, Poland}

\author{P.\,W\"ustner}
\affiliation{Zentralinstitut f\"ur Engineering, Elektronik und Analytik, Forschungszentrum J\"ulich, 52425 J\"ulich, Germany}

\author{M.\,\.{Z}urek}
\thanks{Present address: Argonne National Laboratory, Lemont, IL 60439, USA}
\affiliation{Institut f\"ur Kernphysik, Forschungszentrum J\"ulich, 52425 J\"ulich, Germany}

\collaboration{JEDI collaboration}
	
	
\date {\today}
	

\title{
Supplemental Material: Pilot bunch and co-magnetometry of polarized particles stored in a ring
}
\maketitle

\section{Machine setup}
The typical timing parameters of the setup of the experiment in the accelerator are listed in Table\,\ref{tab:machine-timing}. The beam and machine parameters, as well as the parameters for the operation of the RF Wien filter on flattop, are summarized in  Table\,\ref{tab:machine-conditions}. 
\begin{table}[htb]
	\caption{\label{tab:machine-timing} Timing parameters for the machine operation to provide the beam parameters during the experiments (listed in Table\,\ref{tab:machine-conditions}).}
	\begin{ruledtabular}
	\renewcommand{\arraystretch}{1.2}
		\begin{tabular}{ll}
			Event in cycle   											& Time  [s] \\ \hline
			Injection													& 0 \\
			Acceleration to flattop momentum finished					& 2\\
			Electron cooling on 										& 3\\
			Electron cooling off 										& 50\\
			Electron cooler magnets off									& 55\\
			Carbon target moved in 										& 60\\
			White noise stochastic extraction on 						& 68\\
			Data acquisition on 										& 68\\
			RF Wien filter on 											& 85\\
			White noise stochastic extraction off 						& 174\\
			End of data taking 											& 174 \\
		\end{tabular}
	\end{ruledtabular}
\end{table}
\begin{table}[htb]
	\caption{\label{tab:machine-conditions} Parameters of the deuteron kinematics, the COSY ring, the deuteron elementary quantities, and the field integrals of the RF Wien filter.  The deuteron mass $m$ and the deuteron $g$ factor, taken from the NIST database\,\cite{nist}, are used to specify $G$. The last column indicates when the value is an input (i), calculated (c), or measured (m).}
	\begin{ruledtabular}
	\renewcommand{\arraystretch}{1.2}
		\begin{tabular}{llll}
			Parameter 								& Symbol [Unit] 		& Value & i/c/m\\ \hline
			Deuteron momentum (lab)					& $P$ [MeV/c]			& \num{970.663702}  & c\\
			Deuteron kinetic energy (lab) 			& $T$ [MeV]				& \num{236.284783} & c \\
			Lorentz factor 							& $\gamma$ [1] 			& \num{1.125977} & c\\
			Beam velocity 							& $\beta$  [c] 			& \num{0.459617}  & c\\
			\\
			Nominal COSY orbit circumference 		& $\ell_\text{COSY}$ [m] 			& \num{183.572} & i \\
			Revolution frequency 					& $f_\text{rev}$ [Hz] 	& \num{750602.6} & i\\
			Spin precession frequency $f_s = G \gamma f_\text{rev}$			& $f_s$ [Hz] 			& \num{-120847.303520} & c\\
			\\
			Deuteron mass 							& $m$ [MeV]				& \num{1875.612793}  & i \,\cite{nist}\\
			Deuteron g factor 						& $g$ [1]				& \num{1.714025} & i \,\cite{nist}\\
			Deuteron $G = (g-2)/2$					& $G$ [1]				& \num{-0.142987} & c\\
			\\
			Slip factor								& $\eta$ [1]			& 0.6545 & m\\
			Momentum spread	in middle of cycle		& $\Delta p/p$ [1]		& \num{7.397e-5}  & m\\
			Synchrotron oscillation frequency & $f_\text{sync}$ [Hz] & $205 \pm 21$ & m\\
			\\
			RF Wien filter electric field integral 	& $\int E_x^\text{WF} \dd s$ [V]			&  \num{359.320135} & c \\
			RF Wien filter magnetic field integral  & $\int B_y^\text{WF} \dd s$ [\si{\milli \tesla \meter}]			&  \num{0.0030737339} & c \\
			RF Wien filter active length 					& $\ell_\text{WF}$ [m]	& \num{1.550} & i
		\end{tabular}
	\end{ruledtabular}
\end{table}

\section{Signal and pilot bunch analysis}
The observed longitudinal beam profiles of pilot (p) and signal (s) bunches, shown in Fig.\,1b in the main text exhibit a slight {asymmetry between the center and the head \& tail regions of the bunches. Still their gross features can  be well approximated by Gaussians, 
\begin{equation}
\frac{\dd N_\text{p,s}}{\dd \phi} \propto \exp \left(- \frac{(\phi-\phi_\text{p,s})^2}{2\sigma_\text{p,s}^2}\right)\,.
\label{eq:gauss}
\end{equation}
 At three different times in the cycle, at the start, middle, and end, the parameters of the Gaussians were determined by fitting, and the results are listed in Table\,\ref{tab:two-bunches-para}. Due to the absence of cooling, pilot and signal bunch widths increase by about 60 to 80\% during the cycle. At the $2\sigma$ level, this corresponds to an increasing fractional bunch length with respect to the ring circumference during a cycle from 7 to 13\%.
\begin{table}[htb]
	\caption{\label{tab:two-bunches-para} Results from fits with Gaussians, yielding for pilot (p) and signal (s) bunch the azimuthal locations $\phi_\text{p,s}$ and widths $\sigma_\text{p,s}$  at three times $t_\text{cyc}$ during a typical cycle shown in Fig.\,1b in the main text. The numerical values are expressed in units of radian ($2\pi$ denotes the ring circumference). The last column shows the separation of the signal and pilot bunches, expressed as the ratio of the differences in azimuthal locations to the mean total $4\sigma$-bunch width. Thus, the two bunches are well separated from each other and  in terms of the Gaussian width do not overlap.}
	\begin{ruledtabular}
	\renewcommand{\arraystretch}{1.2} 
		\begin{tabular}{l|ll|ll|c}
			& \multicolumn{2}{c|}{pilot bunch}    & \multicolumn{2}{c|}{signal bunch} &
			bunch separation\\
			Time $t_\text{cyc}$[s] & $\phi_\text{p}$ [rad]& 2$\sigma_\text{p}$ [rad] & $\phi_\text{s}$ [rad] & 2$\sigma_\text{s} $ [rad]
			&  $\frac{\phi_\text{s} - \phi_\text{p}}{2(\sigma_\text{s} + \sigma_\text{p})}$ \\\hline
			\phantom{0}78  & 2.41  & 0.24 & 5.55 & 0.22 & 6.7\\
			122 & 2.43  & 0.35 & 5.58 & 0.36 & 4.4\\
			173 & 2.43  & 0.38 & 5.58 & 0.40 & 4.1\\
		\end{tabular}
	\end{ruledtabular}
\end{table}

\section{RF switches}
\subsection{Switch design}
In order to gate out at least one of the stored beam bunches, high-speed, high-power RF switches have been developed for the driving circuit\,\cite{SlimWFCircuit} of the RF Wien filter\,\cite{Slim2016116}, in close collaboration with the company Barthel\,\footnote{barthel HF–Technik GmbH, 52072 Aachen, Germany \url{https://barthel-hf.com}} that had also built the RF power amplifiers and the components of the driving circuit itself. The switches are characterized by a symmetric switch on and off speed of \SI{20}{ns} and are capable to handle \SI{250}{W} of RF power\,\footnote{It should be noted that such a high switching speed cannot be achieved with a resonant circuit, for instance in an RF solenoid, which is often used as an RF spin flipper in storage rings.}, allowing the Wien filter with four input ports to be operated close to \SI{1}{kW} of total power. The switches are sophisticated, non-commercial, custom-designed active devices, composed of input and output matching circuits, switching High-Electron Mobility Transistors (HMETs), high-current driving circuits of these transistors and high power and heat dissipation systems. 

\subsection{Implementation of the RF switches into the driving circuit}
The modification of the driving circuit of the RF Wien filter (WF) for the present experiment involved the additional implementation of six identical fast RF switches, as indicated in Fig.\,\ref{fig:modified-driving-circuit}. The signal from an RF generator\,\footnote{Model SMB100A, Rohde \& Schwarz GmbH \& Co. KG, Munich, Germany, \url{http://www.rohde-schwarz.de}.} is fed via a 4-way signal splitter into four RF amplifiers\,\footnote{Amplifiers manufactured by Barthel HF-Technik GmbH, Aachen, Germany, \url{http://www.barthel-hf.de}.}  and fed into the waveguide. The purpose of the adjustable elements in the driving circuit allows one to match the device so that the Lorentz force on the orbiting particles can be minimized\,\cite{PhysRevAccelBeams.24.124601}.  Indicated in yellow are the  six fast RF switches. Four switches are installed in each of the four input ports, and two switches are installed at the output of the waveguide of the Wien filter behind the combiners.
\begin{figure}[htb]
	\includegraphics[width=1\columnwidth]{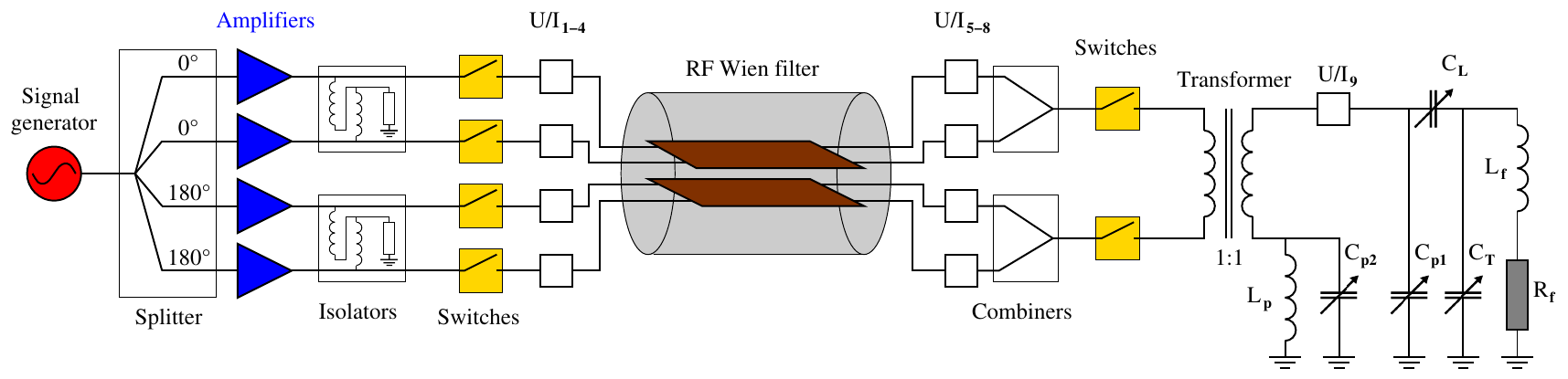}
	\caption{\label{fig:modified-driving-circuit}  Schematic of the driving circuit of the waveguide RF Wien filter, adopted from \,ref.\,\cite[Fig.\,1]{SlimWFCircuit}.}
\end{figure} 

\subsection{Operation of the RF switches during the experiment}
The operation of the RF switches is synchronized with the revolution of the pilot bunches so that the center of the gate is located at $\phi =\phi_\text{p}$ (see Eq.\,(\ref{eq:gauss}) and Table\,\ref{tab:two-bunches-para}). As shown in Fig.\,\ref{fig:Gating}, the gating frequency is equal to the beam revolution frequency $f_\text{rev}=750602.6\, \text{Hz}$, and the temporal gate duration amounts to $T_\text{G} = 0.556\, \mu\text{s}$, while the particle revolution time is $T_\text{rev} = \SI{1.33}{\micro \second}$. The phase gate in Fig.\,\ref{fig:Gating} thus corresponds to $\Delta\phi = 2\pi f_\text{rev}T_\text{G} = 2.62 \, \text{rad}$. Relative to the width $\sigma_\text{p}$ of the beam (listed in Table\,\ref{tab:two-bunches-para}), the gate width $\Delta\phi$ varies from $22\sigma_\text{p}$ at the beginning of the cycle to $14\sigma_\text{p}$ at the end and is thus entirely sufficient for the pilot bunch to pass through the decoupled Wien filter.
\begin{figure}[htb]
	\includegraphics[width=0.5\columnwidth]{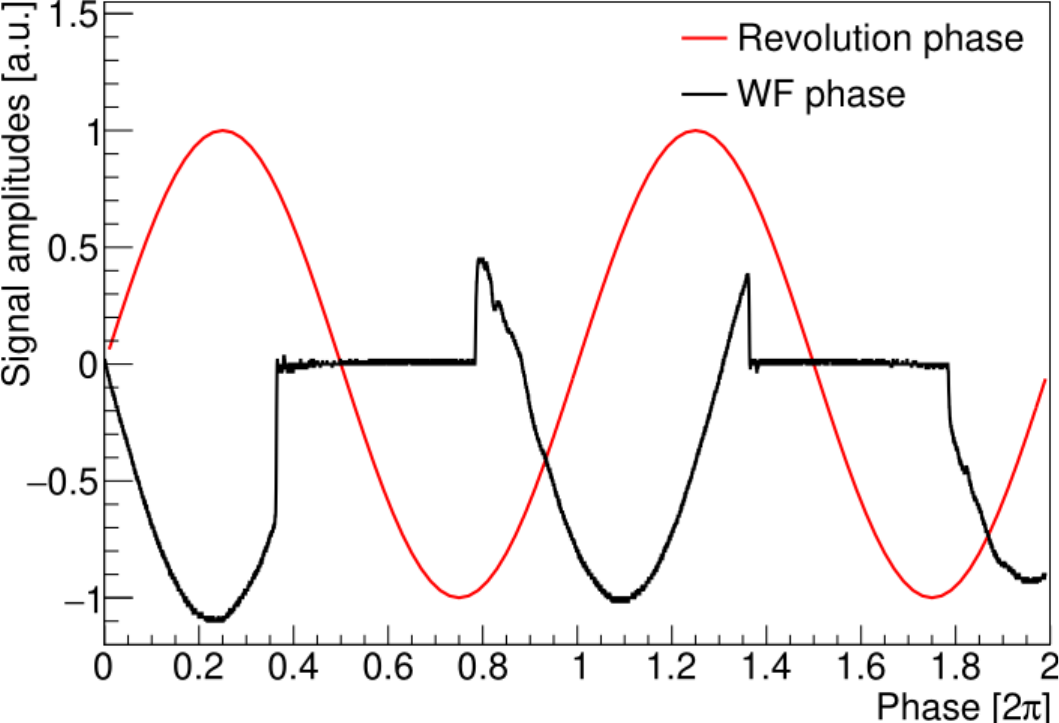}
	\caption{\label{fig:Gating} The signal amplitude of the RF Wien filter, shown in black, is plotted as function of the revolution phase, shown in red, for two beam revolutions in the ring. The signal amplitude of the Wien filter, shown here, was measured during one of the dedicated test runs. The sinus of the revolution phase is shown to emphasize the lock of the gate to the beam revolutions. The variation of the signal amplitude results in a small turn by turn variation of the spin rotation angle in the Wien filter, but this has no adverse effect on the observed spin reversals.}
\end{figure} 

\section{Spin flip analysis of signal and pilot bunches}

\subsection{Signal bunch fit}
The oscillation pattern of the asymmetry in the polarimeter caused by vertical polarization of the  signal bunch, shown in Fig.\,2 of the main text, is fitted with the function $A(t)$, given in Eq.\,(2) of the main text. The fit using data from inside the $\pm 2 \sigma$ boundary of the bunch distribution yields the values listed in Table\,\ref{tab:SF-analysis}.  Our convention is to assign an initial spin-flip phase of zero, resulting in a negative initial spin asymmetry $c$ and implying that the RF Wien filter was effectively switched on at a cycle time of $t_0 \approx \SI{85.5}{s}$.
\begin{table}[htb]
	\caption{\label{tab:SF-analysis} Parameters obtained from a fit of the asymmetry oscillation pattern of the signal bunch, shown in Fig.\,2 of the main text, with Eq.\,(2) of the main text. The $\chi^2/\text{ndf} = 136.071/157 = \num{0.867}$. }
	\begin{ruledtabular}
		\renewcommand{\arraystretch}{1.2} 
		\begin{tabular}{crll}
			Parameter 	   	& Value 			& Error				& Unit 						\\\hline
			$a_\text{s}$       	   	& \num{-4.04}      & \num{0.38} 	    & \SI{e-4}{\per \second} 	\\
			$t_0$		   	& \num{85.548}  	& \num{0.060} 		& \si{\second} 				\\
			$b_\text{s}$				& \num{-0.0228}  	& \num{0.0019}		& 1							\\	
			$c_\text{s}$ 			& \num{-0.0936}	& \num{0.0027}		& 1							\\
			$\Gamma$		&\num{7.30} 	& \num{5.86}	& \SI{e-4}{\per\second}				\\
			$f_\text{SF}$  	& \num{0.07944} 	& \num{0.00010}   	& \si{\hertz}				\\
		\end{tabular}
	\end{ruledtabular}
\end{table}

Making use of the results of the signal bunch fit, listed 
in Table\,\ref{tab:SF-analysis}, the single spin-flip efficiency of the data shown in Fig.\,(2) of the main paper, can be estimated using Eq.\,(3) of main paper, and amounts to 
\begin{equation}
\epsilon_\text{SF} = 1 - \frac{\Gamma}{2 f_\text{SF}} = \num{0.9954} \pm \num{0.0046}\,.
\end{equation}

\subsection{Comparison of center vs head \& tail spin flip frequencies and spin-oscillation amplitudes of the signal bunch}

Particles in the bunch are in constant synchrotron motion. Evidently, only particles with sufficiently large synchrotron amplitudes do contribute to the head \& tail portions of the bunch. They keep oscillating from head to tail and vice versa, spending part of their time also in the central portion of the bunch. Synchrotron oscillations modulate the spin-precession frequency and can affect the spin-flip frequency as well\,\cite{Nikolaev:2023srx}. In order to inspect whether the head \& tail portions of the signal bunches exhibit a noticeable difference regarding the observed spin-flip  frequency, the fit using $A(t)$ was separately applied to data within the head region of the bunch distribution (Set I)  and to the tail region (Set II). For this comparison, only data from within the $\pm 2\sigma_s$ boundary of the signal bunch were used that were split into two statistically \textit{independent} data samples 
\begin{equation}
\begin{split}
	 \text{bunch center} \quad \text{\textbf{Set}\,I:}  \quad & \phi_\text{s} \in [-0.6,+0.6]\sigma_\text{s}\\
	  \text{head \& tail} \quad \text{\textbf{Set}\,II:} \quad & \phi_\text{s} \in [- 2, -0.6]\sigma_\text{s} \, \lor \,  \phi_\text{s} \in [+0.6,+2]\sigma_\text{s}\,,
\end{split}
\label{eq:definition-setsI+II}
\end{equation}	
with about the same number of recorded events. Note that particles with synchrotron amplitudes below $0.6\sigma_\text{s}$ do not contribute to the tail \& head set II. However, in the course of their synchrotron oscillations, particles with larger synchrotron amplitudes populating set II, do spent part of their time in set I   \,\cite{Nikolaev:2023srx}. The results of corresponding fits are presented in Table\,\ref{tab:signal-bunch-SF-analysis}. The observed individual spin-flip frequencies $f_\text{SF}$ of the central and head \& tail regions of the signal bunches are compatible with each other and with the combined results for $f_\text{SF}$ of Sets\,I+II, given in Table\,\ref{tab:SF-analysis}.

The comparison of the properties of the central and head \& tail regions of the signal bunch can be also carried out for the oscillation amplitude $c$ of the two sets I and II, also listed in Table\,\ref{tab:signal-bunch-SF-analysis}. Within the uncertainties, the oscillation amplitudes of the central and the head \& tail regions agree with each other and with the combined value for $c$, listed in Table\,\ref{tab:SF-analysis}.
\begin{table*}[htb]
	\renewcommand{\arraystretch}{1.2}
	\caption{\label{tab:signal-bunch-SF-analysis} Spin flip frequency $f_\text{SF}$ deduced from fits to the vertical asymmetry (see Fig.\,2 in main text) of the signal bunch with  Eq.\,(2) (main text) for different cuts on the longitudinal bunch distribution (see Fig.\,1b in main text). With the boundary between Set I and Set II [defined in Eq.\,(\ref{eq:definition-setsI+II})] at $0.6\sigma_\text{s}$, the two sets yield about the same statistical uncertainties for $f_\text{FS}$.}
	\begin{ruledtabular}
		\begin{tabular}{cccccc}
			Set  &  bunch region & $f_\text{SF}$ [\si{\hertz}] 	& $b$  [1] & $c$ [1] & $\chi^2/\text{ndf}$	\\\hline
			I    & center & $\num{0.07943} \pm \num{0.00014}$ & $\num{-0.006} \pm   \num{0.003}$ & $\num{-0.095} \pm   \num{0.004}$	& $179.16/157 = \num{1.14}$ \\
			II   & head \& tail & $\num{0.07950}    \pm \num{0.00014}$ 		& $\num{-0.027} \pm  \num{0.003}$  & $\num{-0.088} \pm \num{0.004}$ & $132.44/157 = \num{0.84}$\\
		\end{tabular}
	\end{ruledtabular}
\end{table*}

Our analysis suggests, however, one difference between the central (Set\,I) and the head \& tail (Set\,II) regions of the signal bunch in terms of the offset $b$ of the polarimeter asymmetry, listed in Table\,\ref{tab:signal-bunch-SF-analysis}. The fit to the full data within the $2\sigma$ bunch boundary  yielded a value for $b^\text{I+II} = -0.023 \pm 0.002$ (see Table\,\ref{tab:SF-analysis}), while we found for the central part  $b^\text{I} = -0.006 \pm 0.003$ and for the head \& tail regions $b^\text{II} = -0.027 \pm 0.003$. Apparently, within the $2\sigma$ boundary, the  polarimeter asymmetry offset is dominated by interactions of the head \& tail regions of the signal bunch. A possible explanation for this observation might be that head \& tail regions exhibit  larger transverse beam offsets that then generate at the location of the polarimeter target a larger count rate asymmetry compared to target interactions from the central regions of the  bunch. 

\subsection{Determination of the gating efficiency of the RF switches using the pilot bunch}

Albeit being only prototype device that will certainly be improved in future applications, it is interesting to determine from our data on the spin flip frequencies, the efficiency of gating by analyzing the pilot bunch (blue points in Fig.\,2 of the main paper). To this end, we perform a fit using a truncated version of the fit function $A(t)$, given in Eq.\,(2)  in the main paper, where $t_0$ and $f_\text{SF}$ are fixed to the values obtained for the signal bunch (see Table\,\ref{tab:signal-bunch-SF-analysis}). Even before fitting, it is obvious that the oscillation amplitude $c_\text{p}$ is consistent with zero, resulting in vanishing sensitivity to the damping parameter $\Gamma$, so we simply introduced $\Gamma$ as a result of the signal bunch data. The results with the truncated fit function are listed in Table\,\ref{tab:pilot-bunch-with-fixed-t0-and fSF}. 
\begin{table}[htb]
	\caption{\label{tab:pilot-bunch-with-fixed-t0-and fSF} Parameters obtained from a fit of the asymmetry oscillation pattern of the pilot bunch, shown in Fig.\,2 of the main text, with a modified version of Eq.\,(2) of the main text, where the parameters $t_0$,\ $\Gamma$  and $f_\text{SF}$ are fixed to the values obtained from the fit to the signal bunch (see Table\,\ref{tab:signal-bunch-SF-analysis}). The $\chi^2/\text{ndf} = 131.483/160 = \num{0.822}$. }
	\begin{ruledtabular}
		\renewcommand{\arraystretch}{1.2} 
		\begin{tabular}{crll}
			Parameter 	   	& Value 			& Error				& Unit 						\\\hline
			$a_\text{p}$       	   	& \num{-4.10}     & \num{0.36} 	    & \SI{e-4}{\per \second} 	\\
			$b_\text{p}$				& \num{-0.1230}  	& \num{0.0018}		& 1							\\	
			$c_\text{p}$ 			& \num{-0.00074}	& \num{0.00127}		& 1							\\
			\end{tabular}
	\end{ruledtabular}
\end{table}

In terms of the spin flip amplitudes for the signal and pilot bunches  $c_\text{s}$ and $c_\text{p}$, determined form fits to the uncorrelated data sets, the gating efficiency can be expressed as
\begin{equation}
\epsilon_\text{gate} = \frac{c_\text{s} - c_\text{p}}{c_\text{s}} = 1 - \frac{c_\text{p}}{c_\text{s}} = \num{0.9921} \pm \num{0.0135}\,.
\end{equation}
We conclude that within the statistical limits of our experimental test of the pilot beam concept, the gating efficiency is compatible with unity, indicating that our RF switches performed perfectly.

\bibliographystyle{apsrev4-2}
\bibliography{PilotBunch_18.06.2023.bib}
